\documentstyle[aps,multicol,psfig]{revtex}
\begin{document}
\title{Parametric Luminescence of Microcavity Polaritons}
\author{Cristiano Ciuti$^1$, Paolo Schwendimann$^2$ 
and Antonio Quattropani$^1$ }
\address{$^1$Physics Department, Swiss Federal Institute of Technology
Lausanne, CH-1015 Lausanne-EPFL, Switzerland}
\address{$^2$Defense Procurement, System Analysis Division, CH-3003 Bern,
Switzerland}

\date{\today}
\maketitle
\begin{abstract}

The spectral and dispersive emission properties are analytically
determined for the two-dimensional system of exciton-polaritons in
microcavities excited by a resonant and coherent optical pump.  New
collective excitations result from the anomalous coupling between one
generic polariton state and its idler, created by the scattering of
two pumped polaritons. The corresponding parametric correlation is
stimulated by the emitter and idler populations and drives very
efficiently the luminescence.  The intrinsic properties of the
collective excitations determine a peculiar emission pattern.

\bigskip

\end{abstract}
\bigskip

\begin{multicols}{2}

In their pioneering experiments, Weisbuch {\it et al.} \cite{Weisbuch}
discovered a new kind of two-dimensional quasi-particles, resulting
from the strong coupling between quantum well excitons and confined
photons in a semiconductor microcavity.  These peculiar particles,
called microcavity polaritons, have a very sharp dispersion due to the
very light mass of the cavity photon.  Remarkably, this represents a
condensed matter system of small mass quasi-particles which can be
manipulated through laser beams both in frequency and momentum space.
In principle, the low polariton density of states could allow large
occupation numbers at relatively small densities of particles , well
below the critical saturation value due to the fermionic
nonlinearities.  In other words, the polariton system could exhibit
bosonic properties\cite{boson}.  Furthermore, unlike unbound
electron-hole pairs in ordinary semiconductor lasers, microcavity
polaritons have a relatively short time of recombination into external
photons.  All these ingredients are indeed very promising for
applications in the domain of ultrafast all-optical switching and
amplification.

Recently, Dang {\it et al.}\cite{Lesidang} measured photoluminescence
spectra from a II-VI microcavity excited with a {\it nonresonant}
pump.  A threshold was observed in the dependence of the polariton
luminescence intensity as a function of the input power.  Similar
results were reported by Senellart and Bloch in a III-V microcavity
\cite{Senellart}.  These experiments have been interpreted in terms of
enhanced scattering of reservoir excitons into the emitting polariton
modes.  The origin of the enhancement has been attributed to bosonic
stimulation due to final state occupation.

More recently, great insight into the subject has been given by
angle-resolved experiments under {\it resonant} excitation
\cite{Baumberg,Houdre2,Huang,Baumberg2}. 
In this kind of experiments, polaritons are optically excited at a
desired energy and momentum, allowing a direct
control of the polariton dynamics.  In particular, Savvidis {\it et
al.}\cite{Baumberg} have uncovered a new kind of polariton parametric amplifier
through pump-probe experiments.  The
angular selection rules for the parametric conversion of pumped
polaritons into the probe and idler modes are unambiguously given
by the energy and momentum conservation for the polariton scattering.
In the case of an applied probe, the nonlinearity can be explained in
terms of phase-matched wave-mixing of polariton matter beams\cite{PPA}.
Remarkably, giant nonlinearities occur also when the probe beam is
absent and the spontaneous emission is detected
\cite{Baumberg,Houdre2,Baumberg2}.
Spectrally, the luminescence shows a very peculiar distortion with
respect to the bare polariton dispersion\cite{Houdre2,Baumberg2}.
Moreover, the emission is characterized by a finite angular width which
appears to be an intrinsic property of the
system\cite{Houdre2,Baumberg2}.  Indeed, the whole phenomenology
suggests that the interacting polariton matter provides a new kind of
collective excitations.  The nature of these excitations and the
connection with the stimulated scattering mechanism are not yet known.
The explicit solution of this intriguing problem is the motivation and
main result of our present investigation.

In this paper, we present a theoretical analysis of the polariton
nonlinear emission in the case of resonant and coherent pumping.  The
coherence induced by the pump field allows an anomalous coupling
between a generic polariton state and its idler, originated by the
fission of two pumped polaritons.  The anomalous coupling gives rise
to new collective excitations.  The luminescence is driven by the
emitter-idler parametric correlation which is stimulated by the
emitter and idler populations.  We establish how the intrinsic
properties of the collective excitations determine the rich emission
features.  In particular, (i) the peculiar dispersion, (ii) the
lineshape features and (iii) the two-dimensional pattern are provided
by our results.

In order to investigate the nonlinear emission of microcavities in the
strong exciton-photon coupling regime, we work directly in the
polariton basis.  The polariton states have a two-dimensional
character and are described by the in-plane wave-vector and their
spin.  In the following, each ${\bf k}$ will be a two-dimensional
vector in the quantum well plane.  Moreover, we will consider the case
of a circularly polarized pump beam.  Neglecting the spin relaxation,
only the polariton states with a definite circular polarization will
be retained. When showing numerical results, we will use the angular
coordinates $(\theta_x,\theta_y)$ defined by
$(k_x,k_y)=\frac{\omega}{c} (\sin\theta_x,
\sin\theta_y)$, where $\omega$ is the emission frequency.
 Actually, these coordinates represent the angles of the
far-field images in current experiments \cite{Baumberg,Houdre2,Baumberg2}.
Since we consider the
situation of resonant pumping of the lower polariton branch, we can
neglect the nonlinear contribution due to the upper branch.
The destruction operator
 for a lower polariton with in-plane wave-vector ${\bf k}$ is $p_{\bf
 k} = X_{k} b_{\bf k} + C_{k} a_{\bf k}$, where $b_{\bf k}$ and
 $a_{\bf k}$ are the exciton and photon Bose
 operators respectively and $X_{k}$, $C_{k}$ the corresponding
 Hopfield coefficients ($X_{k}>0$ and $C_{k}<0$).  The lower polariton
 Hamiltonian\cite{PPA} is $H=H_{LP}+H^{eff}_{PP}+H_{qm}$.  The free
 term $H_{LP}=
\sum_{{\bf k}} E_{LP}(k)~ p^{\dagger}_{\bf k} p_{\bf k}$ contains
the lower polariton energy dispersion $E_{LP}(k)$. The 
polariton-polariton interaction term reads 
$$ H^{eff}_{PP}=
\frac{1}{2} 
\sum_{{\bf k},{\bf k'},{\bf q}}
\frac{\lambda_X^2}{A} V^{PP}_{{\bf k},{\bf k'},{\bf q}}
~p^{\dagger}_{{\bf k}+{\bf q}}~p^{\dagger}_{{\bf k'}-{\bf q}}
~p_{\bf k}~p_{\bf k'},
$$
where $\lambda_X$ 
is the two-dimensional exciton radius and A is the macroscopic
quantization area.
The interaction potential is $V^{PP}_{{\bf k},{\bf k'},{\bf q}} =
\{ 2 \frac{\hbar \Omega_R}{n_{sat}\lambda_X^2}
(|C_{{\bf k}+{\bf q}}| X_{\bf k'}  
+ |C_{{\bf k'}}| X_{{\bf k}+{\bf q}})
+ \frac{6e^2}{\epsilon \lambda_X} X_{{\bf k}+{\bf q}}X_{{\bf k'}} \}
X_{{\bf k'}-{\bf q}}X_{{\bf k}}
$, with $n_{sat}=7/(16\pi \lambda_X^2)$  the exciton saturation 
density and $\epsilon$ the quantum well dielectric constant.
Notice that $V^{PP}_{{\bf k},{\bf k'},{\bf q}}$ is
positive and represents a repulsive interaction. 
The cavity system interacts with  the external electromagnetic
field through the standard quasi-mode coupling
Hamiltonian $ H_{qm}= \int d\Omega \{
\sum_{{\bf k}}g C_{\bf k}~p^{\dagger}_{{\bf k}}  
~\alpha_{{\bf k},\Omega}
+ H.c.\}~.
$
The operator $\alpha_{{\bf k},\Omega}$ destructs an external
photon with in-plane wave-vector ${\bf k}$ and
frequency $\Omega$.
The spectrum of the polariton luminescence is 
$$
PL({\bf k},t,\omega) \propto |C_{k}|^2 {\mathcal R}e \int_{0}^{+\infty}
 d\tau e^{-i(\omega-i0^+) \tau} 
\langle p^{\dagger}_{\bf k}(t+\tau) p_{\bf k}(t)  \rangle~,
$$
where $p_{\bf k}(t)$ is the time-dependent polariton operator.  Here
and in the following we use the Heisenberg representation of
time-dependent operators and take expectation values on the stationary
state.  In the steady-state regime, the two-time correlation $\langle
p^{\dagger}_{\bf k}(t+\tau) p_{\bf k}(t) \rangle$ depends only on the
relative time $\tau$, and not on $t$.  This implies that the
stationary spectrum can be calculated at a fixed time $t=t_{0}$ when
the steady-state regime is achieved. For simplicity, we choose
$t_{0}=0$ and hence $ PL({\bf k},\omega) \propto |C_{k}|^2 {\mathcal
R}e
\{\langle \tilde{p}^{\dagger}_{\bf k}(\omega) p_{\bf k}(0) \rangle\}$, where 
$\tilde{p}_{\bf k}(\omega)  =
\int_{0}^{+\infty} d\tau e^{i(\omega+i0^+) \tau}p_{\bf k}(\tau) $.

An applied cw-optical pump with in-plane wave-vector ${\bf k_p}$
drives a polariton polarization $\langle p_{\bf k_p}(t)\rangle =
e^{-i\omega_p t} |\langle p_{\bf k_p}\rangle|$, where $\omega_p$ is
the laser frequency resonant with the lower polariton mode.  We
are interested in the generic polariton operator $p_{\bf k}(t)$
with ${\bf k} \neq {\bf k_p}$.  The polariton-polariton interaction
couples $p_{\bf k}(t)$ to the pumped mode and to the idler polariton
operator $p_{{\bf k_{idler}}}(t)$, where ${\bf k_{idler}}=2{\bf
k_p}-{\bf k}$.  This corresponds to the fission process $\{{\bf
k_{p}},{\bf k_{p}}
\}\rightarrow \{{\bf k},{\bf k_{idler}}\}$.
Disregarding the pump noise, we
can perform the replacement $p^{\dagger}_{\bf k}(t)p^{\dagger}_{{\bf
k_{idler}}}(t) p_{\bf k_p}(t) p_{\bf k_p}(t) \simeq 
p^{\dagger}_{\bf k}(t) p^{\dagger}_{{\bf
k_{idler}}}(t) \langle p_{\bf k_p} (t) \rangle^2.
$
Furthermore, we neglect multiple scattering, that is interaction 
between modes other than the 
pumped one (i.e. $\{{\bf k},{\bf k'}
\}\rightarrow \{{\bf k}+{\bf q},{\bf k'}-{\bf q}\}$).
The limit of validity of this approximation will be discussed later on.
With these assumptions, the Heisenberg equations of motion read
 $$
i\hbar
\frac{d}{dt}
\left
(
\begin{array}{c}
p_{\bf k}(t)  \\
p^{\dagger}_{\bf k_{idler}}(t) e^{-i2\omega_p t}\\
\end{array} 
\right
)
=
\hat{M}^{par}_{\bf k}
\left
(
\begin{array}{c}
p_{\bf k}(t)  \\
p^{\dagger}_{\bf k_{idler}}(t) e^{-i2\omega_p t}\\
\end{array} 
\right
)
$$
$$
+
\left
(
\begin{array}{c}
F_{\bf k}(t)  \\
-F^{\dagger}_{\bf k_{idler}}(t) e^{-i2\omega_p t}\\
\end{array} 
\right 
) ~,
$$
where the coupling matrix is 
$$
\hat{M}^{par}_{\bf k}
= 
\left (
\begin{array}{cc}
\tilde{E}_{LP}({\bf k}) -i\gamma_{k} & E^{int}_{\bf k} {\mathcal P}^2_{\bf k_p}
~~\\
- E^{int}_{\bf k} {\mathcal P}^{\star 2}_{\bf k_p}~~
& 2 \hbar \omega_p - \tilde{E}_{LP}({\bf k_{idler}}) -i\gamma_{k_{idler}} 
\end{array}
\right ) ~.
$$
\columnwidth3.4in
\begin{figure}[h!]
\centerline{\psfig{file=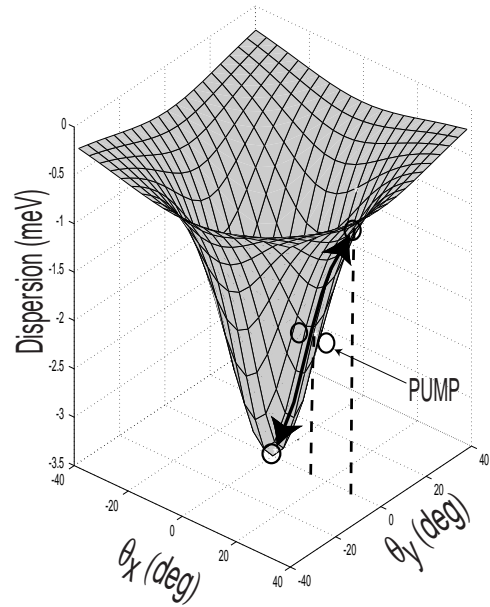,width=2.6in}}
\caption{Lower Polariton energy dispersion $E_{LP}-E_{exc}(0)$
(meV) versus the angles $\theta_x$ and $\theta_y$ (deg).
A sketch is shown for a parametric fission of two pumped polaritons.
}
\label{fig1}
\end{figure}
The off-diagonal term contains the energy $ E^{int}_{\bf
k}=(V^{PP}_{{\bf k_p},{\bf k_{p}},{\bf k-k_p}}+ V^{PP}_{{\bf k_p},{\bf
k_{p}},{\bf k_p- k}})/2$ and the pump-induced polarization
${\mathcal P}_{\bf k_p}= \frac{\lambda_x}{\sqrt{A}} \langle p_{\bf
k_p}\rangle$. We point out that $|{\mathcal P}_{\bf k_p}|^2
=n_{p} \lambda_X^2$
is the {\it coherent density} of pumped polaritons in units of $\lambda_X^{-2}$.
The diagonal process $\{{\bf k},{\bf k_{p}} \}\rightarrow \{{\bf k},{\bf k_{p}}\}$
produces the renormalization of the energy dispersion (blueshift), namely
$\tilde{E}_{LP}({\bf k}) = E_{LP}(k)+ 2 V^{PP}_{{\bf k},{\bf k_p},{\bf
0}} |{\mathcal P}_{\bf k_p}|^{2}$. Notice the equality of the
diagonal elements of $\hat{M}^{par}_{{\bf k}}$ is equivalent to the exact
energy conservation for the process $\{{\bf k_{p}},{\bf k_{p}}
\}\rightarrow \{{\bf k},{\bf k_{idler}}\}$.
Finally,
the coupling to the external photons is
responsible for a radiative damping 
$\gamma_k= 2\pi g^2|C_{k}|^2/\hbar$ and a Langevin
force  
$F_{\bf k}(t) = \int d\Omega 
~g ~C^{\star}_{\bf k} e^{-i\Omega t} \alpha_{{\bf k},\Omega}(0)
$.

From the coupled equations for the operators $p_{\bf k}(t)$ and 
$p^{\dagger}_{\bf k}(t)$, we can obtain the physical quantities of
interest. Let us consider the polariton occupation number
$N_{\bf k}(t) = \langle p^{\dagger}_{\bf k}(t) p_{\bf k}(t) \rangle $.
From the equations for $p_{\bf k}(t)$, we get:  
$$
\frac{d}{dt} N_{\bf k}(t)   
= -\frac{2\gamma_{k}}{\hbar} ~ N_{\bf k}(t) 
+\frac{2}{\hbar}~{\mathcal I}m
\left \{\langle p^{\dagger}_{\bf k}(t) F_{\bf k}(t) \rangle \right.
$$
$$
\left .
+  E^{int}_{\bf k}{\mathcal P}_{\bf k_p}^2
e^{-i2\omega_p t} 
\langle p^{\dagger}_{\bf k}(t) p^{\dagger}_{\bf k_{idler}}(t) \rangle   \right \}~.
$$
This equation shows that the polariton population $N_{\bf k}(t)$
is driven by the {\it anomalous} quantum correlation
$\langle p^{\dagger}_{\bf k}(t) p^{\dagger}_{\bf k_{idler}}(t) \rangle$.
The anomalous correlation evolution reads:
$$
i\hbar \frac{d}{dt} 
\langle p^{\dagger}_{\bf k}(t) p^{\dagger}_{\bf k_{idler}}(t) \rangle =
$$
$$
[-\tilde{E}_{LP}({\bf k})-\tilde{E}_{LP}({\bf k_{idler}})-i(\gamma_k
+\gamma_{k_{idler}})]~
\langle p^{\dagger}_{\bf k}(t) p^{\dagger}_{\bf k_{idler}}(t) \rangle 
$$  
$$
+ \langle p^{\dagger}_{\bf k}(t) F^{\dagger}_{\bf k_{idler}} (t) \rangle
+ \langle F^{\dagger}_{\bf k}(t) p^{\dagger}_{\bf k_{idler}} (t) \rangle  
$$
$$
- E^{int}_{\bf k} {\mathcal P}_{\bf k_p}^{\star 2}
e^{i2\omega_p t} ( 1 + N_{\bf k}(t) + N_{\bf k_{idler}}(t) ) 
)~.$$
Notice that the emitter-idler correlation
$\langle p^{\dagger}_{\bf k}(t) p^{\dagger}_{\bf k_{idler}}(t) 
\rangle
$ has a generation term proportional to the
Boson enhancement factor $( 1 + N_{\bf k}(t) + N_{\bf k_{idler}}(t) )$.
This means that there is a spontaneous {\it inhomogeneous} contribution
and a stimulation term due to the emitter and idler populations.
\columnwidth3.4in
\begin{figure}
\centerline{\psfig{file=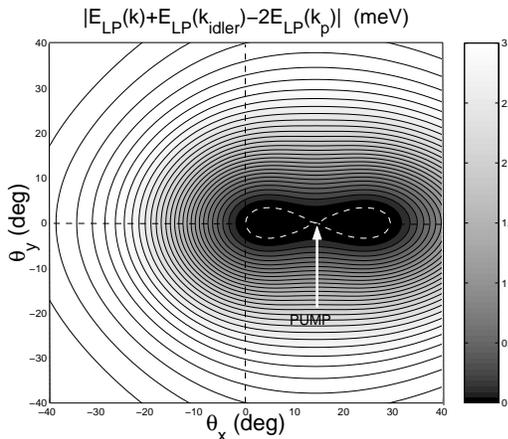,width=2.74in}}
\caption{Contour plot of $|E_{LP}(k)+E_{LP}(k_{idler})-2E_{LP}(k_p)|$
(meV) as a function of ${\bf k} = \frac{\omega}{c} (\sin\theta_x,\sin\theta_y)$.The white dashed line represents the zero value contour,
i.e. exact energy-momentum conservation for the
process $\{{\bf k_{p}},{\bf k_{p}}\} \rightarrow \{{\bf k},{\bf k_{idler}}\}$. 
The x-direction is that of the pump wave-vector ${\bf k_p}$.}
\end{figure}

In order to obtain the frequency spectrum of the parametric luminescence,
we have to come back to the equations for $p_{\bf k}(t)$ and 
$p^{\dagger}_{\bf k}(t)$.
In the frequency space (through
the transformation $\int_{0}^{\infty} e^{i(\omega+i0^+)t} dt$), the
coupled equations read $$
\hbar \omega
\left
(
\begin{array}{c}
\tilde{p}_{\bf k}(\omega)  \\
\tilde{p}^{\dagger}_{\bf k_{idler}}(2\omega_p-\omega) \\
\end{array} 
\right
)
=
\hat{M}^{par}_{\bf k}
\left
(
\begin{array}{c}
\tilde{p}_{\bf k}(\omega)  \\
\tilde{p}^{\dagger}_{\bf k_{idler}}(2\omega_p-\omega) \\
\end{array} 
\right
) 
$$
$$
+\left
(
\begin{array}{c}
\tilde{F}_{\bf k}(\omega) + i\hbar p_{\bf k}(t=0) \\
-\tilde{F}^{\dagger}_{\bf k_{idler}}(2\omega_p-\omega) 
+i\hbar p^{\dagger}_{\bf k_{idler}}(t=0)\\
\end{array} 
\right 
) ~.
$$
Neglecting the noise source $\tilde{F}_{\bf k}(\omega)$
due to the external photons, 
the solution of this linear
inhomogeneous system is 
$$
\tilde{p}_{\bf k}(\omega) =
\frac{
 \Delta(\omega)~i\hbar p_{\bf k}(t=0)
+ E^{int}_{\bf k}{\mathcal P}_{\bf k_p}^2~i\hbar p^{\dagger}_{\bf
k_{idler}}(t=0)  } { (\hbar
\omega - E_{+,{\bf k}})(\hbar \omega -E_{-,{\bf k}})}
~,
$$
with 
$
\Delta(\omega) =
\hbar \omega + \tilde{E}_{LP}({\bf k_{idler}})+ i\gamma_{k_{idler}}-2\hbar\omega_p
$. {\it The complex pole energies $E_{+,{\bf k}}$ and $E_{-,{\bf k}}$
are the eigenvalues of the parametric matrix $\hat{M}_{\bf k}^{par}$
and describe the new collective excitations of the microcavity
system}.  Through the expectation value ${\mathcal R}e
\{\langle \tilde{p}^{\dagger}_{\bf k}(\omega) p_{\bf k}(0) \rangle\}
= {\mathcal R}e
\{\langle  p^{\dagger}_{\bf k}(0) \tilde{p}_{\bf k}(\omega) \rangle\}$, we 
obtain the stationary luminescence as a function of the steady state
population $N_{\bf k}^s =\langle p^{\dagger}_{\bf k}(0) p_{\bf k}(0)
\rangle$ and parametric correlation amplitude
${\mathcal A}_{\bf k}^{s\star}= \langle p^{\dagger}_{\bf k}(0) 
p^{\dagger}_{\bf k_{idler}}(0)\rangle$. The result is
$$
PL({\bf k},\omega) \propto |C_k|^2 {\mathcal R} e \left
\{
i \frac{
\Delta(\omega) N_{\bf k}^{s} + E^{int}_{\bf k} {\mathcal P}_{\bf k_p}^2 {\mathcal A}_{\bf k}^{s \star }
}
{(\hbar \omega - E_{+,\bf k})(\hbar \omega -E_{-,{\bf k}})}
\right \}
.
$$
Simple algebra shows that steady-state solutions of 
the coupled equations for population and parametric correlation are
$\langle p_{\bf k}(t) p_{\bf k_{idler}}(t) \rangle
={\mathcal A}^{s}_{\bf k}~e^{-i2\omega_p t}$ and $N_{\bf k}(t)=N^{s}_{\bf k}$.
The anomalous correlation amplitude reads
$$
{\mathcal A}^{s}_{\bf k} = 
\frac{E^{int}_{\bf k} {\mathcal P}_{\bf k_p}^{2} ~\delta_{\bf k}}
{|\delta_{\bf k}|^2-\frac{(\gamma_k+\gamma_{k_{idler}})^2}{\gamma_k 
\gamma_{idler}} |E^{int}_{\bf k} {\mathcal P}_{\bf k_p}^{2}|^2}
,
$$
 with $\delta_{\bf k}=(2\hbar \omega_p-\tilde{E}_{LP}({\bf
k})-\tilde{E}_{LP}({\bf k_{idler}})-i(\gamma_k +\gamma_{k_{idler}}))$.
Moreover, the steady-state population is 
$$
N^{s}_{\bf k} =
\frac{1}{\gamma_{k}} {\mathcal I}m (E^{int}_{\bf k}{\mathcal
P}_{\bf k_p}^2 {\mathcal A}_{\bf k}^{s\star}).
$$
Of course,  $N^{s}_{\bf k_{idler}}= 
N^s_{\bf k} \gamma_k/\gamma_{k_{idler}}$.
As we will show below, the PL signal (as well as the
parametric correlation amplitude ${\mathcal A}^{s}_{\bf k}$ and population
$N_{\bf k}^s$) diverges when
the pump density $|{\mathcal P}_{\bf k_p}|^{2}$ reaches a threshold
value. In fact, above threshold, the pump polarization can be no longer
treated as a parameter. This means that 
the equation of motion for the pumped mode has to be included,
accounting for the pump depletion. 
Moreover, when the transfer from the pump wave-vector to the other ones
becomes very massive, multiple scattering can be no longer neglected.
Hence, {\it the results here presented
are valid only below the threshold of the parametric luminescence}.
\columnwidth3.4in
\begin{figure}
\centerline{\psfig{file=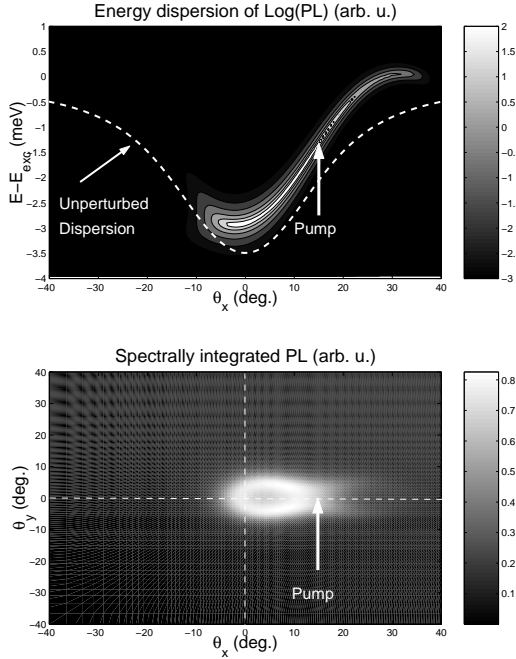,width=2.74in}}
\caption{(a)Contour plot of the parametric photoluminescence (log scale)
as a function of the angle $\theta_x$ (deg)  and energy (meV)
for the pumped density $n_p= 0.03~n_{sat}$ and $\theta_y=0$.
The dashed-line is the unperturbed dispersion of the lower polariton branch.
(b) Two-dimensional pattern of the spectrally integrated emission. 
}
\label{fig3}
\end{figure}
For sake of illustration, we present our analytical results by using
GaAs parameters for a realistic microcavity structure with Rabi
splitting equal to 7 meV. We consider the situation of zero
exciton-photon detuning, i.e. $E_{exc}(0)=E_{cav}(0)$. For simplicity,
we consider a k-independent polariton linewidth $\gamma_k \approx 0.4$
meV (for large k, the decreasing of the radiative width is usually
compensated by nonradiative broadening).  The two-dimensional
dispersion of the lower polariton branch is shown in Fig. 1.  We
consider a pump which resonantly excites the polariton branch at the
critical wave-vector such as $E_{LP}({0})+E_{LP}(2k_p)= 2 E_{LP}(k_p)
$, allowing the particular polariton fission $\{{\bf k_p},{\bf k_p}\}
\rightarrow
\{{\bf 0},{\bf 2k_p}\}$ (with increasing pumping, 
one has to consider the renormalized dispersion $\tilde{E}_{LP}$
instead of $E_{LP}$). Fig. 2 contains the contour plot of the 
quantity $|E_{LP}({\bf
k})+E_{LP}({\bf k_{idler}})- 2 E_{LP}({\bf k_p})|$ in {\bf k}
space (${\bf k_p}$ is along the x-direction).
The white-dashed line represents the zero value contour, i.e. exact
energy conservation for the fission $\{{\bf k_p},{\bf k_p}\}
\rightarrow \{{\bf k},{\bf k_{idler}}\}$. 
Remarkably, the conservation is only
weakly forbidden in a large squeezed portion of ${\bf k}$-space (dark
region) around ${\bf k_{p}}$.  This implies that the parametric
coupling is efficient on a broad angular range.
This is actually
found in Fig.3(a), where a typical contour plot of the photoluminescence
(log scale) is shown as a function of the angle $\theta_x$
($\theta_y=0$) and of the emission energy.  
Moreover, Fig. 3(b) shows the two-dimensional
pattern of the spectrally integrated emission (linear scale).  The
depression of the emission at large angles is due to the vanishing
cavity fraction $|C_{\bf k}|^2$.
\begin{figure}
\centerline{\psfig{file=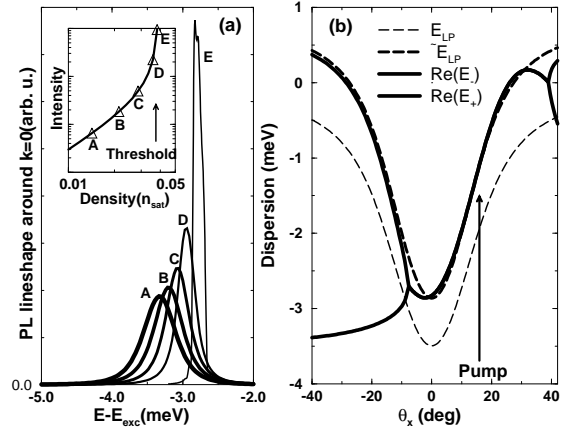,width=2.74in}}
\caption{(a) Normalized emission lineshape around k=0 ($\theta_x=\theta_y=0$;
angular acceptance of 2 deg) for 5 densities (A-E).
Inset: log-log plot of the emission intensity versus  
pumped density ($n_{sat}$ units). 
(b) Dispersions (meV) versus $\theta_{x}$ (deg)
for $\theta_y$ =0 at the highest pump density.
 Thin dashed line: unperturbed polariton dispersion $E_{LP}$.
Thick dashed line: blueshifted dispersion $\tilde{E}_{LP}$.
Thick solid line: ${\mathcal R}e(E_{\pm})$ with 
${\mathcal R}e(E_{-}) \le {\mathcal R}e(E_{+})$.
The bifurcation occurs when the violation of the energy conservation
is large enough and implies negligible parametric emission.
}  
\label{fig4}
\end{figure} 
To determine the dependence on the pump intensity, we just need to
analyze the evolution of the eigenvalues $E_{\pm,{\bf k}}$.  For a
given ${\bf k}$ satisfying exactly the energy-momentum conservation
for the parametric scattering, the expression for the pole energies
$E_{\pm,{\bf k}}$ is very simple, i.e.  ${\mathcal R}e (E_{\pm,\bf k})
= \tilde{E}_{LP}({\bf k})$ and $|{\mathcal I}m (E_{\pm,\bf k})|= \left (
\sqrt{\gamma_k \gamma_{k_{idler}}}
\pm E^{int}_{\bf k} |{\mathcal P}_{\bf k_p}|^{2} \right )
$. This means that the imaginary part of the pole energy $E_{-,{\bf
k}}$ gets smaller at the expense of the other pole $E_{+,{\bf
k}}$. Consequently, the photoluminescence spectrum $PL({\bf
k},\omega)$ gets spectrally narrower.
 The threshold density for the
parametric luminescence is defined by $Im(E_{-,{\bf k}})=0$, i.e.
$E_{int} |{\mathcal P}_{\bf k_p}|^{2} =
\sqrt{\gamma_{k}\gamma_{k_{idler}}}$. For this value, the PL
signal, the parametric correlation amplitude ${\mathcal A}_{\bf k}^s$
and steady-state population $N_{\bf k}^s$ diverge and therefore
the pump depletion has to be included.
All these features can be seen
explicitly in Fig. 4.  With increasing pump density the lineshape of
the emission around $k=0$ blueshifts and narrows (Fig. 4(a)). The
output intensity (see the log-log plot inset) shows a threshold
behavior as a function of the coherent density of pumped polaritons
$n_{p}=|{\mathcal P}_{\bf k_p}|^{2}/\lambda_X^2$.  Notice that when
the polariton damping $\gamma_k$ is small enough, the threshold
density $n_{thr}$ can be orders of magnitude smaller than the exciton
saturation density $n_{sat}$. With the realistic material parameters
employed here, we have $n_{thr}\approx 0.05 n_{sat}$. 
Finally, Fig.4(b), shows explicitly the energy dispersions of the
collective excitations.  When the energy conservation for the
parametric fission is well satisfied, ${\mathcal R} e (E_{\pm}({\bf
k})) \simeq \tilde{E}_{LP}({\bf k})$.  On the other hand, when the
energy conservation is strongly broken and consequently the parametric
luminescence is negligible, ${\mathcal R} e (E_{-}({\bf k}))$ and
${\mathcal R} e (E_{+}({\bf k}))$ bifurcate, tending to the diagonal
elements of the matrix $M_{\bf k}^{par}$, i.e.  $\tilde{E}_{LP}({\bf
k})$ and $2\hbar\omega_p-\tilde{E}_{LP}({\bf k_{idler}})$.  In the
intermediate case, consistently with experiments, the dispersion
${\mathcal R} e (E_{\pm}({\bf k}))$ shows peculiar features such a
change of sign of the in-plane group velocity
\cite{Baumberg2}
around ${\bf 2k_{p}}$ and a
flattening\cite{Houdre2,Baumberg2} around ${\bf k}={\bf 0}$ 
(see also the emission spectra in Fig. 3).

In conclusion, we have theoretically described the parametric
luminescence of microcavity polaritons, showing the subtle interplay
between parametric correlation and stimulated scattering.  The new
collective excitations resulting from the anomalous emitter-idler
coupling produce a very peculiar luminescence pattern in frequency and
momentum space, giving a key to recent experiments
\cite{Baumberg,Houdre2,Baumberg2}. 
Further investigations are in progress for the regime above threshold
not treated in this paper, but investigated in new experiments\cite{above}.
 We wish to thank J.J. Baumberg,
B. Deveaud, R. Houdr\'e, M. Saba, P.G. Savvidis, 
R.P. Stanley, F. Tassone and
C. Weisbuch for fruitful discussions.

\end{multicols}
\end{document}